\documentclass[12pt]{article}
\newcommand{\ba}{\begin{array}}
\newcommand{\ea}{\end{array}}

\newcommand{\be}{\begin{equation}}
\newcommand{\ee}{\end{equation}}
\newcommand{\bea}{\begin{eqnarray}}
\newcommand{\eea}{\end{eqnarray}}
\newcommand{\beal}{\setcounter{letter}{1} \begin{eqnarray}}
\newcommand{\eeal}{\addtocounter{equation}{1} \end{eqnarray}}
\newcommand{\none}{\nonumber \\}
\newcommand{\req}[1]{Eq.(\ref{#1})}
\newcommand{\reqs}[1]{Eqs.(\ref{#1})}
\newcommand{\larrow}{\,\,\,\,\hbox to 30pt{\rightarrowfill}
\,\,\,\,}
\newcommand{\slarrow}{\,\,\,\hbox to 20pt{\rightarrowfill}
\,\,\,}

\newcommand{\half}{{1\over2}}

\newcommand{\bm}{\bibitem}
\begin{document}

\begin{titlepage}
\renewcommand{\thefootnote}{\fnsymbol{footnote}}
\renewcommand{\baselinestretch}{1.3}
\medskip
\hfill  UNB Technical Report 03-xx\\[20pt]

\begin{center}
{\large {\bf Using 3D Stringy Gravity to Understand the Thurston Conjecture}}
 \\ \medskip  {}
\medskip

\renewcommand{\baselinestretch}{1}
{\bf
J. Gegenberg $\dagger$
G. Kunstatter $\sharp$
\\}
\vspace*{0.50cm}
{\sl
$\dagger$ Dept. of Mathematics and Statistics and Department of Physics,
University of New Brunswick\\
Fredericton, New Brunswick, Canada  E3B 5A3\\
{[e-mail: lenin@math.unb.ca]}\\ [5pt]
}
{\sl
$\sharp$ Dept. of Physics and Winnipeg Institute of
Theoretical Physics, University of Winnipeg\\
Winnipeg, Manitoba, Canada R3B 2E9\\
{[e-mail: gabor@theory.uwinnipeg.ca]}\\[5pt]
 }

\end{center}

\renewcommand{\baselinestretch}{1}

\begin{center}
{\bf Abstract}
\end{center}
{\small
We present a string inspired 3D Euclidean field theory as the starting point 
for a modified Ricci flow analysis of the Thurston conjecture. In addition 
to the metric, the theory contains a dilaton, an antisymmetric tensor field 
and a Maxwell-Chern Simons field. For constant dilaton, the theory appears 
to obey a Birkhoff theorem which allows only nine possible classes of 
solutions, depending on the signs of the parameters in the action. Eight 
of these correspond to the eight Thurston geometries, while the ninth 
describes the metric of a squashed three sphere. It therefore appears that 
one can construct modified Ricci flow equations in which the topology 
of the geometry is encoded in the parameters of an underlying field theory. 
}
\vfill
\hfill  June 2003 \\
\end{titlepage}

\section{Introduction:  How to Prove Uniformization Theorems}

The uniformization theorem in two dimensions \cite{poincare} states 
that a closed orientable 
two dimensional manifold with handle number 0,1, $>1$ respectively admits 
{\it uniquely} 
the constant curvature geometry with positive, zero, 
or negative curvatures.  This theorem has proved to be a very powerful tool in 
two-dimensional physics, with applications in conformal field
theories and string theory.  Indeed, in the path-integral
formalism, one must sum over two-dimensional topologies and
geometries.  
Hence, in the path 
integral, one can sum over deformations of each of these geometries, then sum 
over the handle number.  

The potential importance of a 3D uniformization theorem 
is also evident, particularly in the context of 
(super)membrane physics or three-dimensional quantum gravity
where one should be able to perform path-integral quantization via a similar 
procedure to that in two dimensions.
Unfortunately, there is no uniformization theorem in three dimensions, only
a conjecture due to W.P. Thurston. 
\cite{thurston}\footnote{A fairly clear exposition of this can be found in the 
review article by P. Scott \cite{scott}.  There has been some use made of
the Geometrization Conjecture in cosmology, beginning with the paper by
Fagundes \cite{fag}.  A non-exhaustive list of other work along these 
lines is in \cite{kodama, yasuno,weeks,piot,koike}. 
The role of the 
conjecture in high-energy physics is explored in \cite{alvarez}.}.  
This conjecture states that a three-manifold with a given 
{\it topology} has a canonical decomposition into `simple 
three-manifolds,' each of which admits one, and only one, of eight
homogeneous {\it geometries}: $H^3$, $S^3$, $E^3$, $S^2\times S^1$, 
$H^2\times S^1$, Sol, Nil and SL(2,R). The first three geometries are 
maximally symmetric, and hence isotropic.  The remaining five are 
anisotropic, and hence less symmetric, but all have at least a 
three parameter group of isometries.  \footnote{Relativists are familiar 
with the classification of homogeneous geometries into the Bianchi models.  
The relationship of the latter to the Thurston geometries is delineated in 
\cite{fag}.}
The conjecture has not been
completely proven, but considerable progress has been made by Thurston 
\cite{thurston} and recently there has been speculation that Perelman 
\cite{perelman} has overcome some roadblocks in Hamilton's 
program to prove the conjecture using the `Ricci flow' 
\cite{hamilton1,caochow}.   

To prove a uniformization theorem, one must show that a differentiable 
manifold with a given topology admits a certain (highly symmetric) metric.  
This is a formidable task, amounting to showing that structures admitted on 
small part of the manifold can be extended to cover the entirety of it.  An 
alternative is to assume that the manifold admits {\it some} metric, then 
show that the latter can be deformed by some process into the required 
highly symmetric metric.  The obvious choice of the deformation is to 
treat the initial metric akin to a source of heat, and let the heat flow in 
`time', via some parabolic system of partial differential equations.  The idea is that heat 
`uniformizes', so that at the end, as $t\to\infty$, the temperature is 
uniform.

The problem with this strategy is that one must find a parabolic system by 
which to flow the metric.  In other words, the flow should be
\be
\dot g_{\mu\nu}={\cal O}(g_{\mu\nu}),
\ee
where the operator $\cal O$ is elliptic.  Furthermore, this operator must 
transform as a tensor under coordinate transformations.  This restricts 
the operator to be constructed from the curvature tensor.  An obvious 
choice is the Ricci flow \cite{hamilton1}
\be
\dot g_{\mu\nu}=-2 R_{\mu\nu}.
\ee

The first problem with this flow is that it is not strictly parabolic, though 
it shares many of the properties of simpler heat equations.  In fact, it is 
not even a linear partial differential system, since the inverse of the metric 
appears in the Ricci tensor on the right hand side.\footnote{In 
an earlier work, one of us, (J.G.) in collaboration with S. Braham 
\cite{braham} 
considered the Yang-Mills flow as a tool for examining 
uniformization theorems.  Although promising in 2D, the 3D case 
proved intractable.}Thus one must first 
resolve some hard problems in analysis:  the existence of `short-time' flows; 
extension to finite times;  the occurence of singularities {\it before} 
the flow might otherwise uniformize.  Many of these issues have been 
resolved by Hamilton and others \cite{caochow}, but not all.  In fact, a recent 
preprint by Perelman \cite{perelman} claims to have removed one of the 
major impediments in Hamilton's program to prove Thurston's Conjecture via 
a slightly modified version of the Ricci flow.

There is another problem with the Ricci flow:  how do you `input' 
the topology of the manifold?  After all, uniformization really boils 
down to showing that a {\it given} manifold topology implies admissibility of 
a unique (up to diffeomorphisms) homogeneous metric.  For the case 
of closed 2D manifolds, it is 
pretty straightforward to modify the Ricci flow so that topology is 
specified.  This is due to the fact these manifolds are characterized as 
topological spaces by their Euler number $\chi(M_2)$, and the latter is 
related to the curvature via the Gauss-Bonnet formula:
\be
\chi(M_2)={1\over2\pi}\int_{M_2}d^2 x\sqrt{g}R(g).
\ee
The modified Ricci flow \cite{hamilton1} is then
\be
\dot g_{\mu\nu}=-2 R_{\mu\nu}+{\chi(M_2)\over V(M_2)}g_{\mu\nu},
\ee
where $V(M_2):=\int_{M_2}d^2 x \sqrt{g}$ is the volume of $M_2$.  This flow 
has as its fixed points the constant curvature geometries.  It was proved 
by Hamilton \cite{hamilton3} and Chow \cite{chow} that this flow exists, and 
that it converges to its fixed points- thus giving a new proof of the 
two dimensional uniformization theorem.  The important point here is that 
the topology of $M_2$ is explicit data in the second term in the flow.

Unfortunately, the 3D case is not nearly so simple.  Besides the problem that 
in general the 3D versions of the Ricci and modified Ricci flows have 
singularities, the topology of a closed 3D manifold cannot be easily expressed 
as a data point on the right side of a modified Ricci flow. For example, the 
Euler number of a closed 3D manifold is zero.  Thus a term in the 3D 
modified Ricci flow proportional to the metric itself could not be 
obviously specified 
by topological data.

Our proposal here is that if the flow is broadened to include other fields 
defined on 3D space, then one can input at least some of the topological 
data in an explicit way, much as in the 2D case.  As will be shown in the 
next section, such a flow is suggested by the low energy limit 
of a bosonic string propogating in 3D space. In particular, there exists sectors of the theory, labelled by the values (actually signs) of parameters in the Lagrangian, each of which admit one, and only one, of 
the eight Thurston geometries, up to coordinate transformations. We are able to prove uniqueness of the solutions for the case when the Chern-Simons term in 
the theory is turned off; but although plausible, uniqueness in the remaining two sectors has not rigorously been proven. Nonetheless it seems to us that these results are significantly encouraging and suggest that one might be able to use the flow in this model to clarify and understand, and perhaps finally prove, the Thurston Conjecture.

\section{Modified 3D Stringy Gravity}

Of the Thurston spaces, only $E^3$, $S^3$
and $H^3$ are solutions of Einstein gravity with an appropriate 
cosmological constant term. In search of a single theory 
from which all eight of the Thurston geometries arise, we turn to the 
low-energy limit of three-dimensional string theory, which has a metric 
$g_{\mu\nu}$, dilaton $\phi$, Abelian 2-form potential $B_{(2)}$ with
field strength $H_{(3)}=dB_{(2)}$ and a `constant' term in the level of
the original sigma model \cite{3dstring,kal}. This theory has many more 
solutions than the constant curvature geometries.  In fact, it has 
propogating modes.  If the dilaton is set to a constant value, then for 
a given sign for the coupling of the $H^2$ term, the only solutions have 
either constant, non-negative or non-positive metrics.  The value of 
the cosmological constant is given by a constant of integration.
But even with the dilaton non-trivial, the remaining five anisotropic Thurston 
geometries are {\it not} solutions.  

We therefore modify the above 3D stringy theory by appending to it a $U(1)$ gauge field 
with potential 1-form $A$ and field strength $F$ which couple as a
`Maxwell-Chern-Simons theory'.  The 
corresponding action is given by \cite{thurston11d}:
\bea 
S&=&\int d^3x\sqrt{g}\, e^{-2\phi}\left(-\chi+R+4|\nabla\phi|^2 
-{\epsilon_H\over12}H_{\mu \nu \rho}H^{\mu \nu \rho}
-{\epsilon_F\over2}
F_{\mu \nu}F^{\mu \nu} \right)\nonumber\\
& & +{e\over2} 
\epsilon^{\mu\nu\rho}A_{\mu} F_{\nu \rho}, \label{3daction}
\eea
where the last term is the Abelian Chern-Simons term for the one-form potential
$A_{(1)}$, and $F_{(2)}=dA_{(1)}$. 
The Wess-Zumino field $B_{\mu\nu}$ 
is a 2-form potential whose field strength $H_{\mu\nu\rho}=\partial_{[\mu}B_{\nu\rho]}$.  Hence, in 3D, the 
field strength is proportional to the LeviCivita {\it tensor}: 
\be
H_{\mu\nu\rho}=H(x)\eta_{\mu\nu\rho}, \label{1}
\ee
where $H(x)$ is a scalar field.  The equations of motion for the 
`Wess-Zumino field' $B_{\mu\nu}$ are 
\be
H^{\nu\rho}:=\nabla_\mu(e^{-2\phi}H(x)\eta^{\mu\nu\rho})=0.\label{2}
\ee
It is easy to see that the latter implies that $H(x)=c=$ constant.
Without loss of generality, this result can be substituted into the remaining equations 
of motion.  The result is:
\be
E_{\mu\nu}:=R_{\mu\nu}+2 \nabla_\mu\nabla_\nu\phi -{\epsilon_H\over2} 
c^2 e^{4\phi} g_{\mu\nu}-\epsilon_F F_\mu{}^\rho F_{\nu\rho}=0;\label{3}
\ee
\be
J^\mu:=\epsilon_F \nabla_\nu \left(e^{-2\phi}F^{\mu\nu}\right)-{e\over2}\eta^{\mu\nu\rho} F_{\nu\rho}=0;\label{4}
\ee

\be
D:=-\chi+R(g)+  4 \Delta \phi - 4|\nabla\phi|^2 -{\epsilon_H \over2}c^2 e^{4\phi} -{\epsilon_F\over2}F^{\mu\nu}F_{\mu\nu}.\label{5}
\ee

In this paper we will look for solutions with $\phi=0$. In this case, by taking appropriate linear combinations of the trace of (\ref{3}) and  (\ref{5}) one obtains constraints on the Ricci scalar and electromagnetic field strengths.
In particular, the Ricci scalar and the square of the field strength must both be constant:
\be
R=-{\epsilon_H\over2} c^2+2\chi.
\label{Ricci scalar}
\ee
\be
F^{\mu\nu}F_{\mu\nu} = 2\epsilon_F\left(\chi-\epsilon_Hc^2 \right)
\label{F squared}
\ee
These constraints will play an important role in what follows. They have the effect of totally eliminating all propagating modes from the theory, leaving a finite dimensional space of solutions. This can be seen heuristically as follows. The full theory has two coupled dynamical modes, one for the scalar and one for the electromagnetic field. The gravitational field does not propagate in three spacetime dimensions. By restricting to constant $\phi$, the coupling between $\phi$ and $F_{\mu\nu}$ gives rise to the constraint
(\ref{F squared}), which  also eliminates the dynamical mode associated with 
electromagnetic field. Thus, the electromagnetic 
field cannot fluctuate unless the scalar also fluctuates.

 As will be shown below, all of 
the Thurston geometries are solutions 
of the equations of motion of this theory for various values of 
the parameters $\chi,\epsilon_H,\epsilon_F,e$. 
See Table 1 for a list of the solutions.
In particular, the addition of the Maxwell term alone ($e=0$) yields 
$S^2\times E^1$, $H^2\times E^1$ and $Sol$ as solutions. In this case, there exists a generalized Birkhoff theorem which guarantees that these are the only solutions when $\phi=constant$. With $e\neq0$, one finds that the remaining Thurston geometries $Nil$ and $SL(2,R)$ are also solutions. Although it seems plausible that these are the only solutions, we have been unable to find a rigorous proof.

\bigskip
\oddsidemargin -0.5 in
\begin{table}[h]
\begin{center}
\begin{tabular}{||c|c|c|c|c|c|c|c||}
\hline

Geometry & $\epsilon_H$ & $\epsilon_F$ &$\chi$& $e$ &$c$ &$[A_1,A_2,A_3]$ &  Metric\\ \hline \hline

$E^3$ & $*$ &$*$ &$*$ &$*$&$ *$ &   0 &  $dx_1^2+dx_2^2+dy^2$\\ 

$S^3$ & $1$&$ *$&$1$&$ *$&$ 1$&    0 & $ dx_1^2+\sin^2{x_1}dx_2^2+(dy+\cos{x_1}
dx_2)^2$\\ \hline

$H^3$  &  $-1$&$ *$&$-4$&$ *$&$ 2$&  $  0$ &  ${1\over x_1^2}(dx_1^2+dx_2^2+dy^2)$
\\ \hline

$S^2\times E^1$ & $*$&$ 1$&$1$&$ 0$&$ 0$& $[0,\cos{x_1},0]$& $dx_1^2+
\sin^2{x_1}dx_2^2+dy^2$
\\ \hline
$H^2\times E^1$ & $*$&$ -1$&$ -1$&$ 0$&$ 0$ & $[0,{1\over x_1},0]$ & 
${1\over x_1^2}(dx_1^2+dx_2^2)+dy^2$
\\ \hline
$Sol$ &  $-1$&$ 1 $&$2$&$ 0 $&$2$ & $[0,\sqrt{2}x_1,0]$ & $e^{2y}dx_1^2+
e^{-2y}dx_2^2+dy^2$
\\ \hline
$Nil$ &  $1$&$ -1$&$ 0$&$ -1$&$ 1$ & $[0,x_1,0]$ & $dx_1^2+dx_2^2+(dy-x_1 dx_2)^2$
\\ \hline
${\tilde SL(2,R)}$ &$1$&$ -1$&$ -1$&$ 1$&$ 1$ & $[-{\sqrt{2}\over x_2},0,0]$ & 
${1\over x_1^2}
(dx_1^2+dx_2^2)+(dy+ {1\over x_1} dx_2)^2$ \\
\hline
\end{tabular}
\end{center}
\caption{Thurston Geometries as Solutions}
\end{table}
\oddsidemargin -0 in

With the exceptions of Sol and $H^3$, the Thurston spaces can be 
characterized topologically as Seifert fibre bundles $\eta$ over an orbifold 
$Y$.  The topology of a Seifert fibre bundle is determined by the Euler number 
$\chi(Y)$ of $Y$ and the Euler number $e(\eta)$ of the bundle $\eta$ 
\cite{thurston,scott,alvarez}.   Table 2 gives the values of these 
topological invariants for the 3-manifolds which admit Thurston geometries.  
It turns out that for all these spaces, the parameter $\chi$ in the action is related 
to $\chi(Y)$ by $\chi(Y)=\chi$.  In addition, the constant $e$ turns out to be related to
the Euler number $e(\eta)$. 
When the three-form field 
strength $H_{(3)}$ in the string inspired model is real, the parameter $\epsilon_H$ is positive, whereas $\epsilon_H<0$ corresponds to an imaginary$H_{(3)}$. $\epsilon_H$ is positive
only for the Thurston spaces which are Seifert 
fibre bundles, whereas Sol and $H^3$ require $\epsilon_H<0$. 
We shall see below that in all cases where both $\epsilon_F$ and $\epsilon_H$ must be 
specified, they are of opposite sign.

\begin{table}[h]
\begin{center}
\begin{tabular}{||c||c|c|c||}
\hline
 & $\chi(Y)>0$ & $\chi(Y)=0$ &$\chi(Y)<0$ \\  \hline
$e(\eta)=0$ & $S^2\times E^1$ &$E^3$ &$H^2\times E^1$ \\ 
$e(\eta)\neq 0$ & $S^3$&Nil&$\tilde{SL(2,R)}$\\ \hline
\end{tabular}
\end{center}
\caption{Thurston Geometries as Seifert Fibre Spaces}
\end{table}
 
\section{The Electromagnetic Field}
In order to simplify the equations, we define a vector field dual to the Maxwell field strength:
\be
v^\mu:={1\over2}\eta^{\mu\nu\rho}F_{\nu\rho},\label{6}
\ee
where $\eta^{\mu\nu\rho}:=\epsilon^{\mu\nu\rho}/\sqrt{g}$ is the completely skewsymmetric Levi-Civita {\it tensor}.  Then the Maxwell-Chern-Simons  
equation (4) becomes
\be
\epsilon_F\eta^{\mu\nu\rho}\nabla_\nu v_\rho=e v^\mu.\label{7}
\ee
If we multiply by $\eta_{\mu\alpha\beta}$ (contracting on $\mu$) and use the property 
\be
\eta^{\mu\nu\rho}\eta_{\mu\alpha\beta}=\delta^\nu_\alpha \delta^\rho_\beta-\delta^\nu_\beta \delta^\rho_\alpha,
\ee
we get
\be
2\epsilon_F\nabla_{[\alpha}v_{\beta]}=e\eta_{\mu\alpha\beta}v^\mu=e F_{\alpha\beta}.\label{8}
\ee
Now since $F_{\alpha\beta}=2\partial_{[\alpha}A_{\beta]}$, it follows that 
{\it locally} there exists a (piecewise?) smooth function $\sigma$ 
such that
\be
v_\mu=\epsilon_F e A_\mu+\nabla_\mu\sigma.\label{9}
\ee
Globally we would have to append to the above a `harmonic part', determined 
by the topology of the 3-manifold.

Note that $F_\mu{}^\pi F_{\nu\pi}=g_{\mu\nu}v^2-
v_\mu v_\nu$, where $v^2:=g^{\mu\nu}v_\mu v_\nu$. From this \req{F squared} it follows immediately that:
\be
v^2 = {1\over 2}F^{\mu\nu}F_{\mu\nu}=\epsilon_F\left(\chi-\epsilon_H c^2\right).\label{k}
\ee
The gravitational equations now take the simple form:
\be
E_{\mu\nu}=R_{\mu\nu}-{\epsilon_H\over 2}c^2 g_{\mu\nu}-\epsilon_F(v^2g_{\mu\nu} - v_\mu v_\nu)=0;\label{eev}
\ee

>From Eq.(\ref{eev}) it is easy to see that on shell, the Ricci scalar has three constant eigenvalues
\bea
\lambda_1={\epsilon_H c^2\over 2},\\
\lambda_2=\lambda_3=\chi-{\epsilon_H c^2\over 2}.
\eea 
with corresponding eigenvectors $\xi_a^\mu, a=1,2,3$ satisfy 
\bea
\xi_1^\mu=\alpha v^\mu, \\
v_\mu \xi_i^\mu=0,  i=2,3,
\eea
where $\alpha$ is a constant.  The second equations tells us that the 
$\xi_i$ are a linearly independent set of eigenvectors.

>From the eigenvalue equation for $\lambda_1$, namely,
\be
R_{\mu\nu}v^\mu = {\epsilon_H c^2\over 2} v_\nu
\label{eigenvalue}
\ee
one can show that:
\be
\nabla_\mu v^\mu =0.
\label{grad v}
\ee
Simply take the covariant derivative of both sides of (\ref{eigenvalue}), and use the Bianchi identity and the constancy of the Ricci scalar to obtain (\ref{grad v}).

It also follows from the Maxwell-Chern-Simons equations of motion that 
$v^\nu$ is constant norm tangent vector to a geodesic, so that 
\be
v^\nu \nabla_\nu v^\mu=0.
\label{geodesic equation}
\ee
Indeed from the first equality in Eq(\ref{8}), it follows 
that after contracting with $v^\alpha$ we get
\be
v^\alpha\nabla_\alpha v_\beta - v^\alpha\nabla_\beta v_\alpha =0.
\ee
Since $v^\alpha v_\alpha =$constant, it follows that the second term above 
vanishes, and the vanishing of the first term is just the geodesic 
equation.

To summarize, the above analysis shows that the field equations require the dual of the electromagnetic field strength to have constant norm, be divergence free, and obey the geodesic equation. These conditions must hold for all values of the parameters in the Lagrangian. 


\section{Gravitational Equations}
We will now use the vector field $v^\mu$ to specify a local coordinate system in which the metric takes a particularly simple form.  Choose the coordinate system $\{x^1,x^2,y\}$ so 
that 
\be
\left({\partial\over\partial y}\right)^\mu=v^\mu.
\label{coord cond}
\ee
 We will denote the dependence 
of a function $f$ on the $x^j$ by $f(x)$.  Then from the constancy of $v_\mu$ it follows that
\be
g_{33}=v^2,
\ee
where $v^2$ is the constant given by (\ref{k}). 

Without loss of generality we can write the metric as
\be
ds^2=h_{ij}(x,y)dx^i dx^j + v^2( dy+a_i(x) dx^i)^2,\label{h}
\ee
where the `2D metric' $h_{ij}$ depends on {\it all} the coordinates 
$x^1,x^2,y$. However, $A_i$ depend only on the $x^j$. This follows from 
the requirement that $v^\mu$ is tangent to a family of geodesics.  Indeed, using 
$v^\mu\nabla_\mu v^\nu=0$, get $g^{\nu i}\partial_y A_i=0$, which gives the 
desired result due to the invertability of $h^{ij}$.   

It is important to note here that we have used up most of the freedom 
we have to 
choose coordinates. Given a fixed vector field $v^\mu$, the form 
of the metric is preserved by
only two restricted types of coordinate transformations. Firstly:
\be
y \to y + f(x)
\ee
which effects a gauge transformation on the vector field $a_i(x)$: 
\be
a_i(x) \to a_i(x) + \partial_i f(x)
\ee

Secondly, we can do $y$-independent coordinate transformations on the $x^i$. This means that we cannot transform away any $y$-dependence in the metric $h_{ij}$.
 
Note that the form of the metric (\ref{h}) suggests that $v^\mu$ is a Killing vector for the full metric. Indeed a straightforward calculation shows that the $i,j$ components of the Killing equation on $v^\mu$
\be
\nabla_i v_j + \nabla_j v_i = \dot{h}_{ij}
\ee
where we have defined the quantity $\dot h_{ij}:=h_{ij,y}$. 
The other components vanish identically. This shows as one might expect that $v^\mu$ is a Killing vector if and only if $h$ is independent of $y$.

We have yet to impose the condition $\nabla_\mu v^\mu=0$ which is 
equivalent to 
\be
h^{ij}\dot h_{ij}=0,
\label{hdot zero}
\ee
where $i,j,...$ indices are lowered and raised by $h_{ij}$ and its inverse matrix $h^{ij}$. This means of course that $h = \det(h_{ij})$ is independent of $y$.

At this stage we have effectively solved the Maxwell-Chern-Simons equations. The only remaining field equations 
are the Einstein equations (\ref{eev}), which in terms of the $h$ `metric' reduce to
\be
E_{yy}:=-{1\over4}\dot h^{ij}\dot h_{ij}+{v^2\over2} \left(
-\epsilon_H c^2+ e^2 v^4\right)=0;
\label{e33}
\ee
\bea
E_{iy}&:=&\half\left[\nabla_j\dot h_i{}^j-a_j\partial_y\dot h_i{}^j
+\half a_i\dot h^{jk}\dot h_{jk}\right]\nonumber\\
  & &+{v^2\over 2}\left(-\epsilon_H c^2+
e^2 v^4\right) a_i=0;
\label{metric eqn2}
\eea
\be
h^{ij} E_{ij}:={\cal R}(h)+\nabla_k(\dot h^{ki}a_i)+a_k\nabla_i\dot h^{ki}
-a_ka_i\ddot h^{ki}-2\epsilon_F v^2+\left(1-{v^2\over2}h^{ij}a_ia_j\right)
\left(-\epsilon_Hc^2+e^2 v^4\right)=0.
\label{metric eqn3}
\ee
In the above, the covariant derivative $\nabla_j$ is with respect to 
the `metric' $h_{ij}$ and $\dot h_i{}^j:=h^{jk}\dot h_{ik}$, etc.
Note that the first term in \ref{e33} is positive definite, which in turn requires the second term to be negative definite. Moreover, if $v^\mu$ is a Killing vector, so that $\dot{h_ij}=0$, $\epsilon_H$ and $\epsilon_F$ must be of opposite sign.

The considerations so far are completely general. 
In the next section show how the Thurston geometries emerge for restricted values of the parameters in the Lagrangian.

\section{Existence of three Killing Vector fields}
So far, for the case of constant dilaton field, we have not been able to 
prove that the only solutions to the equations 
of motion have metrics that are diffeomorphic to the eight Thurston 
geometries.  However, in the case $e=0$, i.e. no 
Chern-Simons contribution,  one can show that the solutions all admit 
at least three algebraically independent Killing vector fields.  This is 
sufficient to show that the metrics are homogeneous, and hence 
diffeomorphic to one of the Thurston geometries.  To demonstrate this, 
we will use a procedure for examining the integrability of the Killing 
equations developed by Eisenhart \cite{eisenhartcont}, \cite{eisenhartriem} 
and Yano \cite{yano}.

We begin with the set of twelve variables $S:=\{\xi_\mu,\xi_{\mu\nu}\}$.  
We impose on 
these variables the set of nine constraints 
\be
F_0:=\left\{\xi_{\mu\nu}=\xi_{\nu\mu},\partial_\mu(v\cdot\xi)=0\right\}
\ee
where $v\cdot\xi:=g_{\mu\nu} v^\mu\xi^\nu$.

Now the system of partial differential equations that are to be satisfied 
by the variables $S$, subject to the constraints $F_0$, are
\bea
\nabla_\mu\xi_\nu &=& \xi_{\mu\nu};\none
\nabla_\mu\xi_{\nu\rho}&=& R_{\nu\rho\mu\pi}\xi^\pi.\label{lpe}
\eea
The latter set is obtained by differentiating the former and using the 
first set of constraints along with the Ricci and Bianchi identities, as 
shown in \cite{yano}, and also in \cite{wald}, for the case when 
the constraint set includes only the first, enforcing the symmetry of 
$\xi_{\mu\nu}$.  The other constraint in $F_0$ is already integrable. 
Now it is shown in \cite{yano} that the integrability of \req{lpe} is 
the set of equations
\be
F_1:=\left\{ {\cal L}_\xi R_{\mu\nu}=0\right\}.\label{F_1}
\ee
We now show that these integrability conditions are identically satisfied 
as long as $e=0$.

We write the field equations for the metric as
\be
R_{\mu\nu}= a g_{\mu\nu}-\epsilon_F v_\mu v_\nu,\label{eomr}
\ee
where $a:=\epsilon_H c^2/2+\epsilon_F v^2$, with $v^2=$constant.  We also 
use the Maxwell-Chern-Simons equations (\ref{7}).  Then it is straightforward 
to show that 
\be
{\cal L}_\xi R_{\mu\nu}=a {\cal L}_\xi g_{\mu\nu} +v_\mu L_\nu +v_\nu L_\mu,
\label{LR}
\ee
where the vector field $L_\mu$ is defined as
\be
L_\mu:=\nabla_\mu (g_{\nu\rho}\xi^\nu v^\rho)+e \epsilon_F\eta_{\mu\nu\rho}\xi^\nu
v^\rho.\label{L}
\ee
So if $e=0$, both terms in \req{LR} vanish by the constraints $F_0$.

By standard theorems in partial differential equations \cite{eisenhartcont} 
we can express the variables in the set $S$ as a convergent multivariable 
Taylor series 
containing twelve parameters determined from the values of the fields at the 
centre of the Taylor expansion.  But the nine constraints then reduce the 
number of independent parameters to three.  Hence, our fields $\xi_\mu$ are 
a three parameter family of Killing vector fields.  This is the result we 
were after.

For the $e\neq0$ case, it is straightforward to show that if 
$v_\mu$ is a Killing vector field, 
then the metric admits a four parameter family of isometries.  This follows 
from a result in Bona and Coll \cite{bonacoll} that if the Ricci tensor 
satisfies the condition given by the field equation \req{eomr} and in 
addition $v_\mu$ is shear-free, then the metric necessarily admits a 
four-parameter isometry group.  It is easy to see that the condition that 
$v_\mu$ is Killing is equivalent to it being shear-free.  There are in 
fact three solutions with $e\neq 0$ and $v_\mu$ a Killing vector field.  
Two of these are precicely the Thurston geometries Nil and $SL(2,R)$.  The 
third is a twisted line or circle bundle over $S^2$, with isometry group 
the product of $SO(3)$ with the group of translations on the fibre.  So 
locally these geometries are Hopf fibrations of the squashed (topological) three-sphere. This geometry, which forms the boundary of Taub-Nut-AdS and 
Taub-Bolt-AdS space, has recently been studied in the context of the 
holographic principle\cite{squashed}.

So if the theory admits any solutions which are not homogeneous geometries, 
then those geometries have $e\neq 0$ and $v_\mu$ is not a Killing vector 
field.  In the context of the field equations [\ref{hdot zero},\ref{e33}-\ref{metric eqn3}],
this requires solutions with both $\dot{h}_{ij}$ and $a_i$ non-vanishing. The system appears to
be overconstrained, and the only such solution we have found corresponds to $Sol$ for which
$e=0$. 

\section{Letting it Flow}

We will describe here the flow suggested by our three dimensional gravity 
theory.\footnote{The flow equations below, when the field $A_\mu$ is 
turned off, are the RG flow for a non-linear sigma model coupled to the 
target space metric $g_{\mu\nu}$ and two-form potential 
$B_{\mu\nu}$\cite{pol}.  We are 
currently exploring the incorporation of the target space Maxwell-CS 
field in the sigma model.}  The idea is that the right side of the flow has as its zeroes 
the solutions of the equations of motion \reqs{3} to \reqs{5}:
\bea
\dot g_{\mu\nu}&=&-2 
\left[R_{\mu\nu}+2 \nabla_\mu\nabla_\nu\phi +\left( -{\epsilon_H\over4} 
H_{\mu\rho\tau}H_\nu{}^{\rho\tau}-\epsilon_F F_\mu{}^\rho F_{\nu\rho}\right)\right];\label{F3}
\\
\dot B_{\mu\nu}&=&\nabla_\rho\left(e^{-2\phi}H^\rho{}_{\mu\nu}\right);
\label{FB}\\
\dot A_\mu&=&
 \nabla_\nu \left(e^{-2\phi}F_\mu{}^\nu\right)+
{e\epsilon_F\over2}\eta_\mu{}^{\nu\rho} F_{\nu\rho};\label{F4}\\
\dot\phi&=&
-\chi+R(g)+  4 \Delta \phi - 4|\nabla\phi|^2 -{\epsilon_H \over12}H^2 -
{\epsilon_F\over2}F^2.\label{F5}
\eea

One needs to first specify the values of the parameters 
in the set $\{\chi,e,\epsilon_H\}$, according to the topology 
of the 3-manifold.  If the manifold has the 
topology of a Seifert bundle $\eta$ over an orbifold $Y$, we specify 
$\epsilon_H=+1,\chi=\chi(Y), e=e(\eta)$.  If it is not a Seifert 
bundle then $\epsilon_H=-1$.  It is not clear what the topological 
significance is of the parameters $\chi,e$ for this case.  So far, we 
have not found a topological interpretation for the parameter $\epsilon_F$.  
However, we note in Table 1, that $\epsilon_F$ always has the opposite sign 
of either $\epsilon_H$ or $\chi$.  Hence we tentatively suggest 
that $\epsilon_F$ has the opposite sign of $\chi(Y)$, if the manifold is 
a Seifert bundle with $\chi(Y)\neq 0$; otherwise choose its sign to be 
opposite that of $\epsilon_H$. 

Once the parameters (and hence topology) are specified, one begins with an arbitrary configuration of metric, dilaton field, 
2-form potential $B_{\mu\nu}$ and 
U(1) potential $A_\mu$ as initial conditions for the flow equations
(\ref{F3}-\ref{F5}). If the flow is to be useful then in the case where the 
flow is non-singular, the metrtic should reach the appropriate Thurston geometry.

As a consistency check for the program, we consider the case of locally 
homogeneous initial conditions for the flow.   It should be noted
that the Ricci-Hamilton flow 
of locally homogeneous geometries only converges to the fixed points for 
the case of locally homogeneous and {\it isotropic} geometries \cite{isen}.  
The anisotropic geometries become degenerate or singular \cite{isen}.

We shall consider a few of the details for the flow of an initial geometry which is locally
$H^2\times E^1$.  Thus the metric, U(1) gauge field and dilaton are of the form:
\bea
ds^2&=&{\ell^2\over x_1^2}\left( D_1(t) dx_1^2+D_2(t) dx_2^2\right)+E(t) dy^2;\none
A_\mu&=&[0,A(t){\ell\over x_1},0];\none
\phi&=&\phi(t).
\eea
>From the flow of the metric, we find first that the factor $E(t)$ must be 
constant, and hence can be absorbed by rescaling the $y-$ coordinate.  
Second, it
turns out that for any value of the flow parameter $t$, there must be a constant
$\alpha$ such that $D_2(t)=\alpha D_1(t)$.  The constant $\alpha$ can 
be be absorbed by rescaling $x_2$.  Third, the function $A(t)$ in the gauge
potential is frozen by its flow to be a constant $A(t)=a$.  Finally, we calculate
\be
{dD_1\over d\phi}:={\dot D_1(t)\over\dot \phi(t)}=-2{D_1(t)\left( D_1(t)-
a^2\right)\over \left( D_1^2(t)+2 D_1(t)-a^2\right)}.
\ee
The solution is
\be
\phi(D_1)=\phi_0-\half\left\{D_1+\log\left[D_1\left(-
D_1+a^2\right)^{1+a^2}\right]\right\}.
\ee
Hence we find that  $D_1\to a^2$, in the limit $\phi\to\infty$.  Similar
behaviour occurs for the case of the locally homogeneous flow of 
$S^2\times E^1$.  This is already an improvement over the normalized Ricci 
flow (see below), since in the latter, the locally homogeneous flow is singular in 
these cases.

The above calculation suggests that the dilaton field $\phi$, `normalizes' the flow and can in some sense be considered as the physical flow parameter.  If we had
solved the locally homogeneous flows for $D_1, D_2, \phi$ in terms of $t$, we would
have found that the first two do {\it not} converge to a finite value as
$t\to\infty$. Instead, the fields flow to their fixed points as $t\to -\infty$.  In the usual Ricci flow, the locally homogeneous
and isotropic geometries do not converge to their global `round' form in the limit $t\to \infty$. To accomplish
this, the flow is normalized by adding to it a term $2/3 r g_{\mu\nu}$, where $r$ is
the average value of the Ricci scalar over the manifold\cite{hamilton1, isen}.
These considerations suggest
the idea that occurence of singularities in the flow of the metric is tracked by the flow
of the dilaton field, instead of the rather arbitrary parameter $t$.

\section{Conclusion}
The stringy gravity flow described in the last section is a promising approach to proving
the Thurston Geometrization Conjecture.  We support this claim by the following
observations:
\begin{itemize}
\item
It is quite closely related to the Ricci flow and its various modifications considered by
Hamilton, Perelman and others.  Hence the recent progress made by Perelman in
resolving the analytical properties of these flows will almost certainly apply to the flow
described here.
\item
The parameters that appear in the flow are determined by the topology of the 3-
manifold.  This makes it easier to `input' the 3-manifold into the flow at the beginning.
\item
All the Thurston geometries are fixed points of the flow.  Hence we can follow non-
singular flows directly to the Thurston geometries.
\item
The dilaton field in the flow seems to track the singularities in the flow.  This should
streamline the proceedure of performing surgery on the manifolds in regions where
these singularites occur.
\end{itemize}

Finally, and more speculatively, we believe that underlying the stringy flow is a
quantum field theoretic understanding of the Thurston Geometrization Conjecture.  In
particular, if the stringy gravity on the 3-manifolds are the bulk theory, then the sigma
model whose RG flow is the stringy gravity equations of motion is its holographical dual
theory.  Work along these lines is in progress.

\bigskip\noindent
{\bf Acknowledgments}
We wish to acknowledge useful discussions with S. Das, V. Husain, R. B. Mann, C.
Martinez, R. Troncoso, S. Vaidya, J. Vazquez-Poritz and J. Zanelli.  We thank NSERC
for partial funding.

\end{document}